\documentclass{sf2a-conf2014}
\usepackage{graphicx}
\usepackage{hyperref}
\usepackage[]{natbib}  
\usepackage{epstopdf}

\def\BibTeX{{\rm B\kern-.05em{\sc i\kern-.025em b}\kern-.08em
    T\kern-.1667em\lower.7ex\hbox{E}\kern-.125emX}}
\bibpunct{(}{)}{;}{a}{}{,}  %%%%%%%%%%%%%  A&A bibliography style
\begin{document}

\TitreGlobal{SF2A 2014}

%%-----------------------------------------------------------------
%%      the top matter
%%

\title{Surface rotation of solar-like oscillating stars}

\runningtitle{Surface rotation of solar-like oscillating stars}

\author{T. Ceillier}\address{Laboratoire AIM, CEA/DSM/IRFU/SAp - CNRS - Univ. Paris Diderot, Centre de Saclay, 91191 Gif-sur-Yvette Cedex, France}

\author{R.A. Garc\'ia$^1$}

\author{D. Salabert$^1$}

\author{S. Mathur}\address{Space Science Institute, 4750 Walnut Street, Suite 205, Boulder, Colorado 80301 USA}

%% Keep this line, even if the page will be settled afterwards.
\setcounter{page}{237}

%%-----------------------------------------------------------------

\maketitle

%%-----------------------------------------------------------------
%%        The abstract
%% 
%%  Warning!  within the abstract:
%%  - do not use macros. 
%%  - do not use commands like: \cite, \citet, \citep ... etc.

\begin{abstract}
In this work, we use different methods to extract the surface rotation rate of \emph{Kepler} targets showing solar-like oscillations.
\end{abstract}

%% Insert the keywords (to appear in the ADS indexing)
%% Keywords must be separated by a comma

%%-----------------------------------------------------------------

\section{Introduction}
%%---------------------

Rotation is known to modify heavily the structure and evolution of a star, mainly through transport processes linked to meridional circulation. But as a star evolves, its rotation is modified by magnetic breaking and by its expansion during the subgiant and red giant phases. Moreover, it remains difficult to explain the internal rotation profiles derived thanks to asteroseimology \citep[see for instance][]{2013A&A...555A..54C}.

That is why we study the surface rotation rates of \emph{Kepler} solar-like oscillating stars - including Main-Sequence stars, subgiants and red giants - which are good asteroseimic targets. Using two different corrections of Kepler light curves - PDC-MAP \citep{ThompsonRel21} and KADACS \citep{2011MNRAS.414L...6G} - and two different analyses - wavelets decomposition \citep{2010A&A...511A..46M} and autocorrelation function \citep{2013MNRAS.432.1203M}, we derive a reliable surface rotation rate for a large number of these stars.

We also extract photometric levels of activity for these stars and, using the ages derived by \citet{2014ApJS..210....1C}, we are able to better constrain the age-activity- rotation relations for the different categories of stars in our sample.

\section{Methodology}
%%-------------------------
\label{Methodology}
 
In order to get a robust determination of rotation periods, we use two different ways of correcting the data as well as two different ways of getting an estimation of the rotation period. For each star, we use both data corrected using the PDC-MAP pipeline \citep{2012PASP..124.1000S} and data corrected with the KADACS pipeline \citep{2011MNRAS.414L...6G}. For both sets of data, we get an estimation of the rotation period using both a wavelets analysis \citep[see][]{2013AAS...22130105M} and the autocorrelation function \citep[following][]{2013MNRAS.432.1203M}. We then compare the four different results obtained. If they all concur, we select the period obtained as the rotation period with a high confidence. If it is not the case, we flag the rotation period as uncertain. As a last verification, we check visually all the stars for which a rotation period has been derived.

A good proxy of the activity of a star is the variance of the light curve. This measure is strongly linked to the surface rotation rate of the star. Taking this fact into account, we cut the lightcurve into $5 \times P_{rot}$-long parts and calculate the variance of each of these parts, $S_{ph,k=5}$. The average of these $S_{ph,k=5}$ defines the activity index $\langle S_{ph,k=5} \rangle$ of the star. The length of the parts $5 \times P_{rot}$ has been calibrated on the Sun and other well-known stars \citep{2014JSWSC...4A..15M}. This activity index is thus more reliable than other variability indexes -- such as the $R_{var}$ defined by Basri et al. (2011) -- because it is computed based on the rotation period of the star.

The results of this methodology applied on the sample of solar-like oscillating stars on the Main Sequence and the Subgiant phases have been reported in \citet{2014arXiv1403.7155G}.

\section{Example for a red giant star}
%%-------------------------
\label{RG}

The same methodology has been also applied to a comprehensive sample of \emph{Kepler} Red Giants and will be described in details in Ceillier et al. 2015 (in preparation). Due to their low activity levels, these stars are not supposed to show clear rotational modulations in their lightcurve. This is why only a small fraction of the global sample (around 2\%) give conclusive results. These peculiar Red Giants could result from mergers, as discussed by Tayar et al. 2015 (in preparation).

\begin{figure}[ht!]
 \centering
 \includegraphics[width=0.8\textwidth,clip]{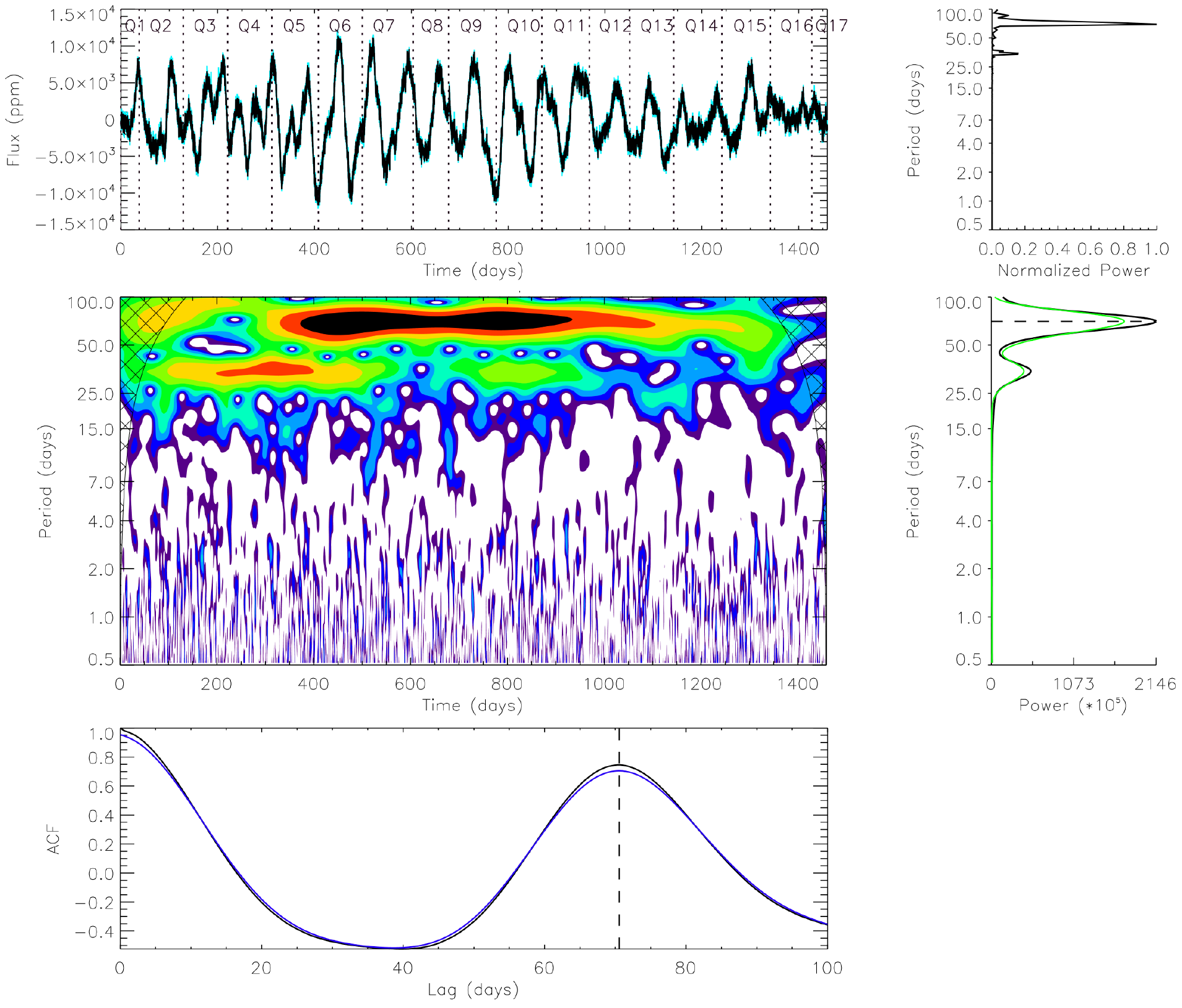}      
%% Note the ABSENCE of the extension .pdf , .eps or .ps  !
  \caption{Example of the analysis for KIC~2570214 (KADAC data). Top panel: Long-cadence \emph{Kepler} light curve (cyan) and rebinned light curve (black), where vertical dotted lines indicate the transitions between the observing quarters. Top right panel: associated power density spectrum as a function of period between 0.5 and 100 days. Middle left panel: Wavelet Power Spectrum (WPS) computed using a Morlet wavelet between 0.5 and 100 days on a logarithmic scale. The black-crossed area is the cone of influence corresponding to the unreliable results. Middle right panel: Global Wavelet Power Spectrum (GWPS) as a function of the period of the wavelet and the associated fit composed from several gaussian functions (thin green line). The horizontal dashed line designates the position of the retrieved $P_{\rm{rot}}$. Bottom panel: AutoCorrelation Function (ACF) of the full light curve plotted between 0 and 100 days (black) and smoothed ACF (blue). The vertical dashed line indicates the returned  $P_{\rm{rot}}$ for the ACF analysis.}
  \label{ceillier:fig1}
\end{figure}

An example of a Reg Giant star's light curve analysis can be seen in Fig.~\ref{ceillier:fig1}. The 70~days modulation is clearly visible.

\section{Conclusions}
%%--------------------
\label{Conclu}

We have now powerful and reliable methods to extract surface rotation rates from \emph{Kepler} light curves. It is thus possible to derive surface rotation rates for a huge number of stars, spread over the whole HR diagram. While the results for dwarves \citep[see][]{2014arXiv1403.7155G} confirm for field stars the age-rotation relation know as the Skumanish law \citep{1972ApJ...171..565S}, the discovery of rapidly rotating, highly active red giant stars opens a new perspective for the evolution of these evolved stars.

% Optional acknowledgements
% -------------------------
\begin{acknowledgements}
SM acknowledges the support of the NASA grant NNX12AE17G. DS, RAG, SM, and TC received funding from the CNES GOLF and CoRoT grants at CEA. RAG also acknowledges the ANR (Agence Nationale de la Recherche, France) program IDEE (n¡ ANR-12-BS05-0008) ``Interaction Des \'Etoiles et des Exoplan\`etes''.
\end{acknowledgements}

%% The following lines are required when using BibTEX (strongly encouraged!):
\bibliographystyle{aa}  % A&A bibliography style file (aa.bst)
\bibliography{ceillier.bib} % your references in file: Yourfile.bib

\begin{thebibliography}{11}
\expandafter\ifx\csname natexlab\endcsname\relax\def\natexlab#1{#1}\fi

\bibitem[{Ceillier {et~al.}(2013)Ceillier, Eggenberger, Garc\'{\i}a, \&
  Mathis}]{2013A&A...555A..54C}
Ceillier, T., Eggenberger, P., Garc\'{\i}a, R.~A., \& Mathis, S. 2013,
  Astronomy and Astrophysics, 555, A54

\bibitem[{Chaplin {et~al.}(2014)Chaplin, Basu, Huber, Serenelli, Casagrande,
  {Silva Aguirre}, Ball, Creevey, Gizon, Handberg, Karoff, Lutz, Marques,
  Miglio, Stello, Suran, Pricopi, Metcalfe, Monteiro,
  Molenda-$\backslash$.Zakowicz, Appourchaux, Christensen-Dalsgaard, Elsworth,
  Garc\'{\i}a, Houdek, Kjeldsen, Bonanno, Campante, Corsaro, Gaulme, Hekker,
  Mathur, Mosser, R\'{e}gulo, \& Salabert}]{2014ApJS..210....1C}
Chaplin, W.~J., Basu, S., Huber, D., {et~al.} 2014, Astrophysical Journal,
  Supplement, 210, 1

\bibitem[{Garc\'{\i}a {et~al.}(2014)Garc\'{\i}a, Ceillier, Salabert, Mathur,
  van Saders, Pinsonneault, Ballot, Beck, Bloemen, Campante, Davies, {do
  Nascimento Jr.}, Mathis, Metcalfe, Nielsen, Suarez, Chaplin, Jimenez, \&
  Karoff}]{2014arXiv1403.7155G}
Garc\'{\i}a, R.~A., Ceillier, T., Salabert, D., {et~al.} 2014, ArXiv e-prints

\bibitem[{Garc\'{\i}a {et~al.}(2011)Garc\'{\i}a, Hekker, Stello,
  Guti\'{e}rrez-Soto, Handberg, Huber, Karoff, Uytterhoeven, Appourchaux,
  Chaplin, Elsworth, Mathur, Ballot, Christensen-Dalsgaard, Gilliland, Houdek,
  Jenkins, Kjeldsen, McCauliff, Metcalfe, Middour, Molenda-Zakowicz, Monteiro,
  Smith, \& Thompson}]{2011MNRAS.414L...6G}
Garc\'{\i}a, R.~A., Hekker, S., Stello, D., {et~al.} 2011, Monthly Notices of
  the RAS, 414, L6

\bibitem[{Mathur {et~al.}(2013)Mathur, Garcia, Metcalfe, Pinsonneault, \& van
  Saders}]{2013AAS...22130105M}
Mathur, S., Garcia, R.~A., Metcalfe, T.~S., Pinsonneault, M.~H., \& van Saders,
  J. 2013, in American Astronomical Society Meeting Abstracts, Vol. 221,
  American Astronomical Society Meeting Abstracts \#221, \#301.05

\bibitem[{Mathur {et~al.}(2010)Mathur, Garc\'{\i}a, R\'{e}gulo, Creevey,
  Ballot, Salabert, Arentoft, Quirion, Chaplin, \&
  Kjeldsen}]{2010A&A...511A..46M}
Mathur, S., Garc\'{\i}a, R.~A., R\'{e}gulo, C., {et~al.} 2010, Astronomy and
  Astrophysics, 511, A46

\bibitem[{Mathur {et~al.}(2014)Mathur, Salabert, Garc\'{\i}a, \&
  Ceillier}]{2014JSWSC...4A..15M}
Mathur, S., Salabert, D., Garc\'{\i}a, R.~A., \& Ceillier, T. 2014, Journal of
  Space Weather and Space Climate, 4, A15

\bibitem[{McQuillan {et~al.}(2013)McQuillan, Aigrain, \&
  Mazeh}]{2013MNRAS.432.1203M}
McQuillan, A., Aigrain, S., \& Mazeh, T. 2013, Monthly Notices of the RAS, 432,
  1203

\bibitem[{Skumanich(1972)}]{1972ApJ...171..565S}
Skumanich, A. 1972, Astrophysical Journal, 171, 565

\bibitem[{Smith {et~al.}(2012)Smith, Stumpe, {Van Cleve}, Jenkins, Barclay,
  Fanelli, Girouard, Kolodziejczak, McCauliff, Morris, \&
  Twicken}]{2012PASP..124.1000S}
Smith, J.~C., Stumpe, M.~C., {Van Cleve}, J.~E., {et~al.} 2012, Publications of
  the ASP, 124, 1000

\bibitem[{Thompson {et~al.}(2013)Thompson, Christiansen, Jenkins, Caldwell,
  Barclay, Bryson, Burke, Campbell, Catanzarite, Clarke, Coughlin, Girouard,
  Haas, Ibrahim, Klaus, Kolodziejczak, Li, McCauliff, Morris, Mullally,
  Quintana, Rowe, Sabale, Seader, Smith, Still, Tenenbaum, Twicken, \&
  Uddin}]{ThompsonRel21}
Thompson, S.~E., Christiansen, J.~L., Jenkins, J.~M., {et~al.} 2013, {Kepler
  Data Release 21 Notes (KSCI-19061-001)}, Kepler mission

\end{thebibliography}

\end{document}